\documentclass[journal,twoside]{IEEEtran}
\usepackage{hyperref}
\usepackage[utf8]{inputenc}

\usepackage{amsmath,amssymb,amsfonts}
\usepackage{bbm}

\usepackage{graphicx}
\usepackage{booktabs}
\usepackage{tabularx}
\usepackage{multirow}
\usepackage{enumitem}
\usepackage[normalem]{ulem}
\usepackage{xcolor}

\usepackage{cite}

\usepackage{balance}
\usepackage{comment}
\usepackage{acronym}

\newcommand{\noteb}[1]{\textcolor{black}{#1}}

\definecolor{gencolor}{RGB}{0,0,0}
\newcommand{\gen}[1]{\textcolor{gencolor}{#1}}
\newenvironment{genblock}{\color{gencolor}}{}
\newcommand{\ours}{\textsc{Textraction}}

\usepackage{tikz}
\usetikzlibrary{positioning,shapes.multipart,shapes.geometric,arrows}

\tikzset{
  block/.style={
    rectangle, rounded corners,
    draw=black, very thick,
    minimum height=3em, minimum width=3em
  },
  input/.style={
    circle, draw=black, very thick,
    minimum size=2em
  },
  output/.style={
    circle, draw=black, very thick,
    minimum size=2em
  }
}

\newcommand{\norm}[1]{\left\lVert#1\right\rVert}

\acrodef{STFT}{short-time Fourier transform}
\acrodef{ISTFT}{inverse short-time Fourier transform}
\acrodef{BSS}{blind source separation}
\acrodef{DOA}{direction of arrival}
\acrodef{DC}{deep clustering}
\acrodef{DPRNN}{dual-path recurrent neural network}
\acrodef{TF}{time-frequency}
\acrodef{TCN}{temporal convolutional network}
\acrodef{ATF}{acoustic transfer function}
\acrodef{SI-SDR}{scale-invariant signal-to-distortion ratio}
\acrodef{MSE}{mean square error}
\acrodef{DFT}{discrete Fourier transform }
\acrodef{SIR}{signal-to-interference ratio}
\acrodef{OVA}{overlap-and-add}
\acrodef{SDR}{signal-to-distortion ratio}
\acrodef{BLSTM}{Bidirectional Long Short-Term Memory}
\acrodef{SOTA}{state-of-the-art}
\acrodef{RI}{Real-Imaginary}
\acrodef{RIR}{room impulse response}
\acrodef{SNR}{signal-to-noise ratio}
\acrodef{RNN}{Recurrent Neural Networks}
\acrodef{FLOP}{floating point operation}
\acrodef{WPE}{weighted prediction error}
\acrodef{STOA}{state of the art}
\acrodef{FC}{fully connected}
\acrodef{E2E}{end-to-end}
\acrodef{SDR}{signal to distortion ration}
\acrodef{SIR}{signal-to-interference ratio}
\acrodef{STOI}{short-time objective intelligibility}
\acrodef{PESQ}{perceptual evaluation of speech quality}
\acrodef{CNN}{convolutional neural network}
\acrodef{MVDR}{minimum variance distortion beamformer}
\acrodef{RTF}{relative transfer function}
\acrodef{IPD}{inter-channel phase difference}
\acrodef{LPS}{logarithm power spectra}
\acrodef{AF}{angle features}
\acrodef{DSB}{delay-and-sum beamforme}
\acrodef{SI-SDR}{scale-invariant signal-to-distortion ratio}
\acrodef{FiLM}{feature-wise linear modulation}
\acrodef{ULA}{uniform linear array}
\acrodef{TPD}{target phase differenc}
\acrodef{SISDR}{scale-invariant signal-to-distortion ratio}
\acrodef{TSE}{Target speaker extraction}
\acrodef{PReLU}{Parametric Rectified Linear Unit}
\acrodef{RPC}{Relative Position Cues}
\acrodef{FiLM}{feature-wise linear modulation}

\acrodef{SC}{single-channel}
\acrodef{MC}{multi-channel}
\acrodef{gLN}{global layer normalization}
\acrodef{CLAP}{Contrastive Language-Audio Pretraining}

\title{Textraction: Unified Text-Guided Speaker Extraction and Enhancement for Single- and Multi-Channel Audio\thanks{This work was supported by...}
}

\author{Aviad Eisenberg,
Sharon Gannot~\IEEEmembership{Fellow,~IEEE,} and Shlomo E.~Chazan
\thanks{\sloppy A. Eisenberg is with Bar-Ilan University, Israel, and with OriginAI, Israel, \texttt{aviad.eisenberg@biu.ac.il}; S. Gannot is with Bar-Ilan University, Israel, e-mail: \texttt{sharon.gannot@biu.ac.il}; and S.~E.~Chazan is with OriginAI, e-mail: Israel, \texttt{shlomi@originai.co}.
}
}

\begin{document}
\maketitle

\begin{abstract}
We present \textsc{Textraction}, a text-guided speaker extraction framework that enables the selective extraction of desired speakers or noise sources from an audio mixture using only natural-language descriptions. The proposed method operates within a single unified model that supports both single-channel and multi-channel inputs and it allows conditioning on either semantic textual descriptions of the target source or on its spatial description.
\gen{Dedicated single- and multi-channel encoders feed a shared stack of time--frequency transformer blocks, which is conditioned on the prompt and on the recording configuration in every block through cross-attention and estimates the target directly from the mixture in one forward pass, without a separation stage or a source-selection stage.}
The system can extract an arbitrary subset of entities from a mixture, assuming up to two simultaneously active speakers and two noise sources, and interprets prompts specifying which entities to \emph{keep}, which to \emph{remove}, or both. In addition, we show that the framework naturally applies to a speech-enhancement setting, in which a single target speaker is present alongside background noise. To circumvent the application of computationally costly large language models used in prior work, we introduce a lightweight, task-adapted text encoder obtained by contrastively fine-tuning a compact text-retrieval model, which learns to be sensitive to instruction polarity (e.g., \emph{keep} vs.\ \emph{remove}) while remaining robust to paraphrasing.
Experiments across attribute-based extraction, spatially guided extraction, and speech enhancement show that the proposed approach effectively isolates the desired entities under various acoustic and spatial conditions\gen{, reaching 11.11~dB (multi-channel) and 10.83~dB (single-channel) \ac{SI-SDR} in attribute-based extraction, 10.53~dB in spatially guided extraction, and 18.66~dB (multi-channel) and 18.15~dB (single-channel) in enhancement,} and matches or surpasses dedicated spatial-extraction and enhancement baselines while offering substantially greater flexibility. Audio samples are available on our demo page.\footnote{\href{https://aviad3189.github.io/Textraction/}{https://aviad3189.github.io/Textraction/}}
\end{abstract}
\begin{IEEEkeywords}
Speaker extraction; Speech enhancement; Multi-microphone processing; Text-guided conditioning; Text embedding
\end{IEEEkeywords}

\section{Introduction}

Humans can effortlessly attend to a desired speaker in the presence of competing talkers and background noise, a capability often referred to as the \emph{cocktail-party effect}. Reproducing this ability computationally is central to many speech technologies. Classical \emph{speech separation} aims to decompose a mixture into all constituent sources, typically without specifying a preferred source, whereas \emph{speaker (or target) extraction} focuses on isolating a particular speaker of interest while suppressing all other sources. Speaker extraction is especially appealing in user-driven scenarios (e.g., teleconferencing, hearing-assistive devices, and voice-controlled systems), where the user either implicitly or explicitly specifies which speaker to enhance.


The literature contains extensive work on both single-channel and multi-channel separation and extraction. Single-channel approaches have achieved remarkable performance with deep learning architectures that learn to disentangle sources from monaural mixtures, often under permutation-invariant training or related objectives \cite{wang2023tf,li2024spmamba,liu2023x,jiang2025dual}. In parallel, multi-channel methods exploit spatial diversity offered by microphone arrays, incorporating spatial features or beamforming-inspired modules to improve robustness in reverberant and noisy environments \cite{bohlender2023spatially,quan2024spatialnet,taherian2024multi,Cohen2024Explainable}. For target speaker extraction \cite{zmolikova2023neural}, a common paradigm is to condition the model on an auxiliary cue that identifies the target, such as a short enrollment utterance \cite{eisenberg2023two,serre2025mtse,Eisenberg2022Extraction}, visual information \cite{pan2022usev,lin2023av,tao2025audio}, or spatial cues derived from an array geometry \cite{tesch2023multi,ge2022spex,zorilua2021investigation,jing2025end,Ellinson2026,Ellinson2026a}. While effective, these cues are not always available in practice: enrollment audio may be missing, cameras may not exist, and multi-channel recordings are not guaranteed.

Recently, text-guided conditioning has emerged as a flexible and powerful paradigm for speech and audio processing, enabling semantic control over complex acoustic tasks. Instead of relying solely on acoustic references, text prompts provide a compact and expressive mechanism for specifying target attributes, speakers, or sound events.

Early works explored text guidance for sound extraction rather than explicit speaker extraction. For example, Yuan et al. \cite{yuan2025flowsep} proposed extracting sounds using text prompts, where a frozen Flan-T5 model was employed to generate text embeddings. Jiang et al. \cite{jiang2024target} applied text prompts in a diarization setting; however, the method only indicates when a speaker is active and does not perform speech extraction.

Several studies have investigated textual cues tailored to specific application scenarios. In \cite{jiang2025tpeech}, dialogue history was used as a textual cue for human–AI question answering. In \cite{jiang2024ptse}, slide or poster content was used for text enrollment without temporal alignment with the audio, targeting meetings, presentations, lectures, and poster sessions. Both studies adopted a pretrained Contrastive Language–Audio Pretraining (CLAP) model \cite{elizalde2023clap} as the text encoder. Similarly, Rahimi et al. \cite{rahimi2022reading} assumed that the exact linguistic content of the target speaker’s utterance is known a priori and used it as enrollment information.

More recently, several text-guided speaker extraction methods have been proposed for single-channel scenarios \cite{hao2025typing, huo2025beyond, shi2025sam, dai2025inter, jiang2024listen, rahimi2025say}. Hao et al. \cite{hao2025typing} introduced a free-text-guided framework in which the desired speaker could be specified via spectral features, textual transcription, or an auxiliary waveform. However, \gen{its mixtures contain only two speakers, without environmental noise sources}. Jiang et al. \cite{jiang2024listen} \gen{considered mixtures of two speech and two non-speech sources, but did not explicitly simulate or evaluate controlled reverberant conditions}. Dai et al. \cite{dai2025inter} proposed the use of relative textual references (e.g., “older,” “second,” “louder”) to identify the target speaker.

The aforementioned text-guided methods primarily operate in single-channel settings, relying on spectral or semantic cues for target speaker identification. A notable exception is the work of Zhu et al. \cite{zhu2025ssdq}, who proposed a multi-channel target speaker extraction framework guided by both semantic and spatial queries. In this framework, spatial queries specify predefined absolute regions in the microphone-array coordinate system, while semantic text queries, embedded with a frozen CLAP text encoder, describe speaker attributes and coarse spatial locations. However, the framework does not model relative spatial relations between speakers.

Despite these advances, several limitations remain. First, no existing text-guided speaker extraction model can seamlessly operate in \emph{both} single-channel and multi-channel scenarios within a unified framework. Second, most prior approaches rely on large pretrained language models for text embedding, which are computationally expensive and typically require parameter-efficient fine-tuning strategies such as LoRA. Conversely, approaches using frozen text encoders lack flexibility in handling diverse prompt formulations, such as specifying entities to retain, remove, or both.

In this work, we propose a unified text-guided extraction framework for both single-channel and multi-channel scenarios. Natural-language prompts specify targets through spectral cues, spatial cues, or their combinations. The framework is robust to high reverberation and multiple noise sources and can extract any entity in the mixture, including noise components, under arbitrary permutations. Free-form prompts allow users to specify which entities to preserve, suppress, or both. We train a lightweight text encoder tailored to the relevant prompt types, improving efficiency and task alignment without relying on large language models. \gen{One network, conditioned on the prompt, estimates the target from the mixture in a single forward pass, and no separation or selection stage is used.}

Our main contributions are summarized as follows:
1) We propose a unified model that supports both single-channel and multi-channel text-guided extraction, effectively leveraging spectral and spatial information within a single framework\gen{. SC and MC inputs enter through dedicated encoders and share all subsequent blocks; the prompt and an SC/MC embedding condition every block, and the network maps the mixture to the target in one pass, without a separate-then-select pipeline.}
2) We introduce a lightweight text encoder based on text retrieval models, which is significantly more efficient than large language models while remaining expressive enough for flexible prompt-based control.
3) For spatial extraction, we incorporate relative position cues \cite{eisenberg2026spatially} expressed in free-form text, enabling the model to exploit relative spatial relationships between speakers.

\section{Problem Formulation}
A scenario with $Q$ concurrently active speakers and $R$ directional noise sources, recorded by an array of $J$ microphones in a reverberant and noisy environment, is considered. The problem is formulated in the \ac{STFT} domain, where $k \in \{0,\ldots,K-1\}$ and $t \in \{0,\ldots,T-1\}$ denote the frequency-bin and time-frame indices, respectively, and $K$ and $T$ are the total numbers of frequency bins and time frames.
The microphone-array observation are then given by:
\begin{equation}
\mathbf{x}(t,k)
= \sum_{q=1}^{Q} s_q(t,k)\,\mathbf{h}_q(k)
+ \sum_{r=1}^{R} n_r(t,k)\,\mathbf{h}_r(k),
\end{equation}
where $s_q(t,k)$ is the clean (anechoic) speech signal of the $q$-th speaker, $\mathbf{h}_q(k)\in\mathbb{C}^{J\times 1}$ is the \ac{ATF} vector between the $q$-th speaker and the microphone array, $n_r(t,k)$ is the $r$-th noise source, and $\mathbf{h}_r(k)\in\mathbb{C}^{J\times 1}$ is the corresponding noise \ac{ATF} vector.

We consider two main operating modes. For the \emph{extraction} task, we focus on mixtures that contain two concurrently active speakers and two directional noise sources, i.e., $Q=R=2$. For the \emph{enhancement} task, we consider a single target speaker, $Q=1$, in the presence of either one or two directional noise sources, i.e., $R\in\{1,2\}$. Both tasks are studied under single- and multi-microphone recording setups. Specifically, we consider a single-channel configuration with $J=1$, as well as a two-microphone configuration with $J=2$, in which spatial information provided by the microphone array is also exploited.

Given the observation $\mathbf{x}(t,k)$ and a natural-language prompt $p$, the goal is to estimate the reference-microphone signal of the entity (or set of entities) described by $p$, whether that entity is a speaker or a directional noise source.

\section{Prompt Generation}
\label{sec:prompts}

\subsection{Attribute-Based Extraction}

For extraction based on semantic attributes, we randomly select the desired and interfering entities in each mixture. Each configuration is represented by a binary selection vector
\[
\mathbf{m} = [m_{s_1}, m_{s_2}, m_{n_1}, m_{n_2}],
\]
where $m=1$ denotes an entity to be preserved and $m=0$ denotes an entity to be removed. For example, the vector $[1,0,1,0]$ corresponds to keeping the first speaker and the first noise source while removing the second speaker and the second noise source.

For each selection vector, we construct three types of \emph{base prompts}: (i) prompts that contain only entities to be preserved, (ii) prompts that contain only entities to be removed, and (iii) prompts that contain both preserved and removed entities. Noise descriptions are taken directly from the dataset metadata. Noise sources whose descriptions include references to human speakers (e.g., \emph{person}, \emph{people}, \emph{speech}) are excluded to avoid semantic ambiguity. Speaker descriptions are defined using their attributes, including gender, relative loudness in the mixture, or a combination of both.

When speaker separation is required (i.e., configurations of the form $[1,0,*,*]$ or $[0,1,*,*]$), the two speakers must differ in at least one attribute. If both speakers share the same gender, one of them is randomly assigned to be louder than the other. This attribute is stored and later enforced during signal generation to ensure consistency between the audio mixture and the textual prompt.

An example of a base prompt is:
\begin{quote}
\emph{``Remove the loud man, keep the weak woman, remove the barking dog, keep the sound of the engine.''}
\end{quote}

To increase linguistic variability, the order of the prompt components is randomly shuffled. Each base prompt is then expanded using a large language model (LLM). In this work, we employ \emph{Gemma~3}~\cite{team2025gemma}, which is prompted to generate five semantically equivalent paraphrases for each base prompt. Since this procedure is applied independently to each of the three base-prompt types, a total of 15 distinct prompts are generated for each mixture configuration.

For each signal, we additionally generate corresponding prompts describing the interfering entities using the same procedure. \gen{We refer to such a prompt as the \emph{complementary prompt}: its target is the complementary set of entities, given by the selection vector $\mathbf{1}-\mathbf{m}$.} During training, the model alternates between conditioning on prompts describing the desired entities and prompts describing the interference.

\subsection{Spatially Guided Extraction}

For extraction based on spatial attributes, the model is trained to extract a single desired speaker from the mixture. Following~\cite{eisenberg2026spatially}, the spatial descriptions are limited to a finite set of predefined textual options, which are summarized in Table~\ref{table:spatial_prompts}. These descriptions express \emph{relative} spatial relationships between the two speakers---for example, which one is closer or more to the right---rather than absolute coordinates in the array frame. To enable free-text conditioning, we employ a large language model to generate $1{,}000$ paraphrased prompts for each spatial description. During training, one prompt is randomly selected from the corresponding set. \gen{Since the target in this task is a single speaker without the noises, the complementary prompt of a spatial scene describes the \emph{other speaker} (e.g., the speaker on the left instead of the speaker on the right), not the complementary set of all four entities.}

\begin{table*}[!ht]
\caption{Base spatial prompts describing speaker position and orientation. Each base prompt is expanded into $1{,}000$ free-form paraphrases.}
\label{table:spatial_prompts}
\centering
\normalsize
\begin{tabular}{|c|l|}
\hline
\textbf{\#} & \textbf{Base Spatial Prompts} \\ \hline
1 & The speaker positioned on the right \\ \hline
2 & The speaker positioned on the left \\ \hline
3 & The speaker located farther away \\ \hline
4 & The speaker located closer \\ \hline
5 & The speaker who is closer and oriented more to the left \\ \hline
6 & The speaker who is closer and oriented more to the right \\ \hline
7 & The speaker who is farther away and oriented more to the left \\ \hline
8 & The speaker who is farther away and oriented more to the right \\ \hline
\end{tabular}
\end{table*}

\subsection{Speech Enhancement}

For the speech enhancement task, where a single target speaker is present alongside background noise, we again use an LLM to generate textual prompts. Specifically, the following base sentences are used: \emph{``keep the talking speaker''}, \emph{``remove the background noise''}, and \emph{``keep the talking speaker and remove the background noise''}. Each sentence is rephrased $1{,}000$ times, and during training a single prompt is randomly selected, following the same procedure as in the spatially guided scenario. \gen{Complementary prompts are used only in the extraction tasks.}

\section{Proposed Model}

\begin{figure*}[t]
    \centering
    \includegraphics[width=\textwidth]{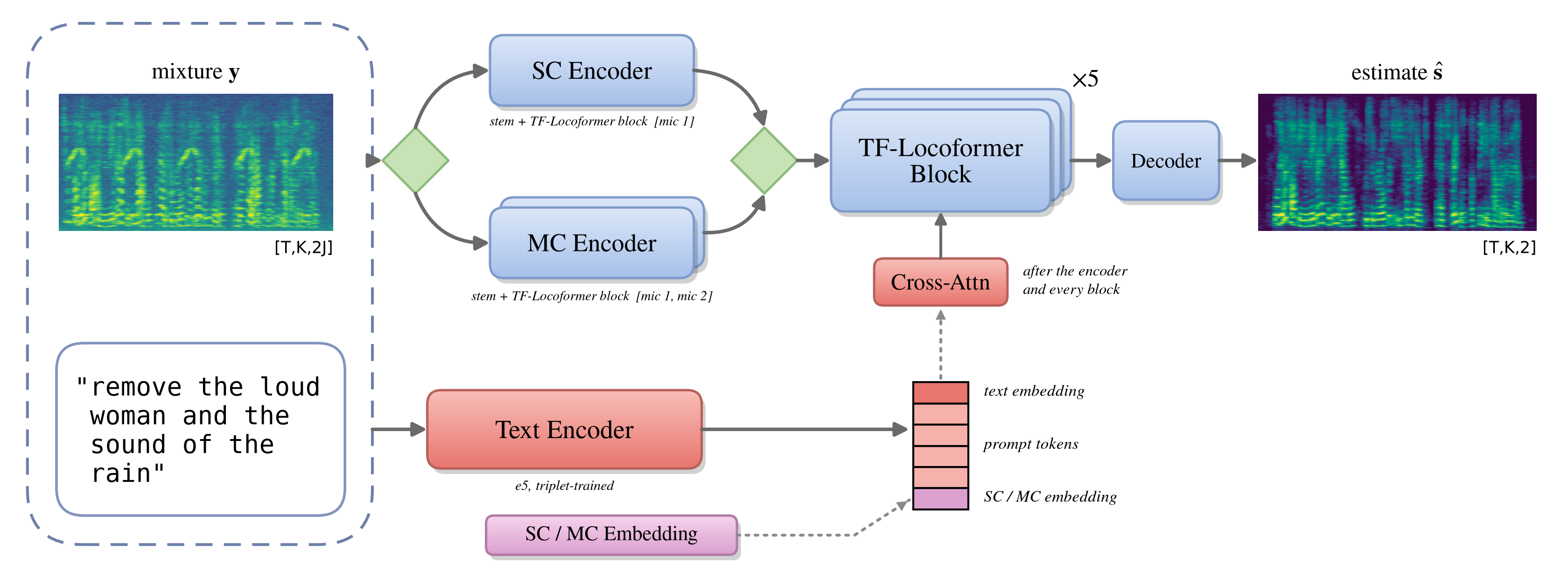}
    \caption{\gen{Overview of the proposed \ours{} model. The mixture $\mathbf{y}$ is represented in the STFT real--imaginary domain and routed, according to the recording configuration, to the SC encoder (reference microphone) or to the MC encoder (both microphones); each encoder is a convolutional layer followed by a TF-Locoformer block. Five shared TF-Locoformer blocks follow. After the encoder and after every shared block, a cross-attention layer conditions the features on the pooled prompt embedding and the token-level states of the fine-tuned lightweight text encoder, together with a learned SC/MC embedding. A decoder outputs the target estimate $\hat{\mathbf{s}}$.}}
    \label{fig:model}
\end{figure*}

\begin{genblock}
Figure~\ref{fig:model} presents an overview of the proposed \ours{} framework. The network receives the mixture and the prompt and estimates, in one forward pass, the signal described by the prompt. It is built from TF-Locoformer blocks~\cite{saijo2024tf}. To enable unified operation in both single-channel (SC) and multi-channel (MC) settings, the input is processed by the corresponding SC or MC encoder, while the remaining blocks are shared between the two configurations. The prompt and a learned SC/MC embedding, which informs the shared network of the active recording configuration, condition the features after the encoder and after every shared block through cross-attention. The network has no separation stage and no source-selection stage; the prompt alone determines which signal is extracted. The following subsections describe the prompt representation and the network architecture in detail.
\end{genblock}

\subsection{Prompt Embedding}

To guide the model toward the desired output, we condition it on free-form textual prompts. Existing approaches often rely on large language models (LLMs) as text embedders; however, such models are computationally heavy and may be unnecessary for the limited semantic scope required in our task. Since the prompts mainly describe source attributes and extraction instructions, we hypothesize that a lightweight model capable of capturing linguistic meaning is sufficient.

Accordingly, we adopt the E5 text-retrieval model~\cite{wang2022e5} as our text encoder. Given an input prompt, the E5 model produces a fixed-dimensional embedding vector. However, embeddings produced by a generic retrieval model are not guaranteed to reflect the semantic distinctions required for our task. For instance, prompts such as \emph{``remove the barking dog''} and \emph{``keep the barking dog''} may be mapped to nearby locations in the embedding space despite conveying opposite instructions.

To address this limitation, we fine-tune the E5 model using task-specific contrastive supervision. For each prompt in our dataset, we construct a triplet consisting of an anchor prompt, a positive prompt, and a negative prompt. These prompts are generated using the \emph{Gemma~3} LLM. Given an anchor prompt, the positive example is a paraphrase with equivalent semantic meaning, generated with as little lexical overlap as possible. The negative example conveys the opposite instruction while preserving similar wording to the anchor.

For example, given the anchor prompt \emph{``Extract the loud man and the audience's applause, remove the woman and the running engine,''} a positive example is \emph{``Focus on the louder male speaker and the sound of a crowd's cheers, while filtering out the female presence and the motor noise,''} and a negative example is \emph{``Remove the loud man and the audience's applause, extract the woman and the running engine.''}

We fine-tune the E5 model using a triplet loss objective, encouraging high cosine similarity between the anchor and positive embeddings while pushing the anchor and negative embeddings farther apart. This training explicitly enforces sensitivity to instruction polarity (e.g., \emph{keep} versus \emph{remove}) while maintaining semantic robustness to paraphrasing.

Table~\ref{tab:e5_scores} presents the cosine similarity scores before and after fine-tuning. While the pre-trained E5 model assigns similarly high similarity scores to both positive and negative prompts, the fine-tuned model successfully distinguishes between them, demonstrating its suitability for text-guided extraction.

\begin{table}[!htbp]
\centering
\normalsize
\caption{Cosine similarity between anchor prompts and their positive and negative counterparts, before and after fine-tuning the E5 model. After fine-tuning, opposite-polarity prompts are pushed apart while paraphrases remain close.}
\label{tab:e5_scores}
\begin{tabular}{lcc}
\hline
 & \textbf{Positive} & \textbf{Negative} \\
\hline
\textbf{E5 (pre-trained)} & 0.9565 & 0.9889 \\
\textbf{E5 (fine-tuned)}     & 0.9911 & $-0.5405$ \\
\hline
\end{tabular}
\end{table}

\begin{genblock}
The triplet-trained E5 model initializes the text encoder of \ours{}, and it is fine-tuned jointly with the extraction network at a low learning rate. The extraction network receives two representations of the prompt. The first is the pooled E5 embedding $\mathbf{e}\in\mathbb{R}^{d}$, where $d$ denotes the embedding dimension: the L2-normalized average of the token states, which is the representation shaped by the triplet training. The second is the sequence of token-level hidden states of E5, which lets the network attend to individual words of the prompt.
\end{genblock}

\subsection{Architecture}
\label{subsec:architecture}

\begin{genblock}
The proposed \ac{SC}--\ac{MC} text-guided extraction network operates in the \ac{STFT} domain using a real--imaginary (RI) representation. Given a mixture tensor $\mathbf{X}\in\mathbb{R}^{B\times C\times F\times T}$, where $B$ is the batch size, $C$ is the number of input channels ($C{=}2$ for the RI parts of the reference microphone and $C{=}4$ for the RI parts of both microphones), and $F,T$ are the frequency-bin and time-frame dimensions, the model outputs an estimate $\widehat{\mathbf{S}}\in\mathbb{R}^{B\times 2\times F\times T}$ of the target in RI form, which is inverted by the inverse \ac{STFT} to give the waveform estimate $\hat{s}$.

\emph{SC and MC encoders.} To support both recording configurations within one model, we use two encoders of identical topology that differ only in their input dimensionality: an SC encoder operating on $C{=}2$ channels and an MC encoder operating on $C{=}4$ channels. Each encoder consists of a 2-D convolutional layer, which maps every time--frequency bin to an embedding of dimension $D$, followed by group normalization and one TF-Locoformer block~\cite{saijo2024tf}. A binary indicator selects the encoder per example: every example passes only through the encoder of its own configuration, and the encoder outputs are merged back into one batch. The MC encoder can therefore exploit the inter-microphone phase and level differences, while SC examples are processed without padding the missing microphone.

\emph{Shared blocks.} The encoder output is processed by five shared TF-Locoformer blocks. Each block alternates between modeling along the frequency axis and along the time axis, and it combines multi-head self-attention with convolutional feed-forward modules, which capture both local and global context. A transposed-convolution decoder maps the final features to the two RI channels of the target.

\emph{Conditioning.} The conditioning sequence $\mathbf{Z}$ consists of the projected pooled prompt embedding, the E5 token states, and a learned SC/MC embedding that indicates whether the current example contains multi-channel spatial information; all its elements have dimension $d$ and are layer-normalized. The features are conditioned after the encoder and after each of the five shared blocks, at six locations in total. At each location, a cross-attention layer lets every time--frequency bin of the mixture feature map $\mathbf{H}\in\mathbb{R}^{D\times F\times T}$, i.e., the output of the encoder or of the preceding shared block, act as a query that attends to the conditioning sequence $\mathbf{Z}$ (the keys and values), and the result is added residually to $\mathbf{H}$. Rather than explicitly defining how the prompt and the recording configuration interact with the audio, the attention mechanism learns these relationships. Because the conditioning enters every block, the network decides at every depth which parts of the mixture to keep, and no explicit separation into sources is needed.

\emph{Initialization.} The encoders, the shared TF-Locoformer blocks, and the decoder are initialized from a four-source TF-Locoformer separator that was trained with permutation-invariant training on the same data: the block of each encoder from the separator's first block, and the five shared blocks from its remaining blocks. The separator is used only for initialization and is not a stage of the model. The decoder is initialized from the separator's decoder by summing its weights over the four sources, the second-microphone input channels of the MC encoder are initialized to zero, and the output layers of all cross-attention layers are initialized to zero, so that at initialization the conditioning layers are the identity.
\end{genblock}

\subsection{Objective Functions}
\label{subsec:objective}
\begin{genblock}
We represent all signals by the real and imaginary (RI) parts of their \ac{STFT}, which serve both as the network input and as the training target. The time-domain estimate $\hat{s}$ is obtained by applying the inverse \ac{STFT} to the network output. The training objective has three terms: an \ac{SI-SDR} term, a log-magnitude term, and a \ac{CLAP} term. The main criterion is the \ac{SI-SDR} between the target signal $s$ and the estimate $\hat{s}$:
\end{genblock}
\begin{equation}
\mathrm{SI\text{-}SDR}(s,\hat{s}) = 10\log_{10}\frac{\norm{\alpha s}^2}{\norm{\alpha s-\hat{s}}^2},
\qquad
\alpha = \frac{\langle \hat{s}, s\rangle}{\norm{s}^2}.
\end{equation}
\begin{genblock}
Here $\alpha s$ is the projection of $\hat{s}$ onto $s$, so the scale of $\hat{s}$ does not change the measure. The loss is the negative \ac{SI-SDR} improvement over the unprocessed mixture $y$ at the reference microphone:
\begin{equation}
\mathcal{L}_{\mathrm{SI\text{-}SDR}} = \mathrm{SI\text{-}SDR}(s,y) - \mathrm{SI\text{-}SDR}(s,\hat{s}).
\end{equation}
The first term does not depend on the network, so it has no effect on the gradients. It only shifts the loss so that $\mathcal{L}_{\mathrm{SI\text{-}SDR}}=0$ means no improvement over the mixture.
\end{genblock}

To encourage accurate spectral envelopes and stabilize optimization, we add an MSE loss on the log-magnitude spectra,
\begin{equation}
\mathcal{L}_{\log|\cdot|} =
\left\|
\log_{10}\!\left(|\widehat{\mathbf{S}}|+\epsilon\right) -
\log_{10}\!\left(|\mathbf{S}|+\epsilon\right)
\right\|_2^2 ,
\end{equation}
where $\epsilon=e^{-6}$.

In examples whose prompt keeps exactly one noise source, we additionally enforce consistency between the estimate and the textual description of that noise. We feed the estimated waveform $\hat{s}$, resampled to 48~kHz, to the audio encoder of a pretrained, frozen \ac{CLAP} model and the dataset caption $c$ of the kept noise clip to its text encoder, obtaining the embeddings $\mathbf{z}_{aud}(\hat{s})$ and $\mathbf{z}_{txt}(c)$, and maximize their cosine similarity:
\begin{equation}
\mathcal{L}_{\mathrm{CLAP}} = -
\frac{\langle \mathbf{z}_{aud}(\hat{s}), \mathbf{z}_{txt}(c)\rangle}
{\|\mathbf{z}_{aud}(\hat{s})\|\;\|\mathbf{z}_{txt}(c)\|}.
\end{equation}
The full training loss is
\begin{equation}
\mathcal{L} = \mathcal{L}_{\mathrm{SI\text{-}SDR}} + \mathcal{L}_{\log|\cdot|} + \mathbbm{1}_{\mathrm{noise}}\,\mathcal{L}_{\mathrm{CLAP}},
\end{equation}
where $\mathbbm{1}_{\mathrm{noise}}$ equals $1$ when the target is a single noise source and $0$ otherwise. \ac{STOI} is used only as an evaluation metric.

\begin{genblock}
\emph{Prompt pairing.} In the extraction tasks, every mixture appears in the batch twice, once with its prompt and once with its complementary prompt (Section~\ref{sec:prompts}). The two examples have identical audio and different targets, which forces the network to rely on the text in order to reduce the loss.
\end{genblock}


\section{Experimental Setup}

\subsection{Dataset}
We generate a simulated text-guided extraction dataset composed of mixtures containing speech and environmental noise. For each mixture, we randomly select two speech utterances from LibriSpeech and two noise clips from VGGSound. To avoid semantic ambiguity between ``noise'' and ``speech,'' we exclude VGGSound samples whose captions indicate human speech activity (e.g., \emph{person talking}, \emph{people speaking}).

Reverberant multi-microphone recordings are synthesized using a room-simulation framework based on the image method~\cite{habets2006room}. For each simulated room, the microphone-array position and its azimuth (with respect to the wall) are randomly selected, while enforcing a minimum distance of $0.5$~m from the nearest wall. The reverberation time ($T_{60}$) is uniformly sampled in the range $0.3$--$0.8$~s. We use a two-microphone array with $8$~cm spacing. The two speakers and the selected noise sources are placed at random positions in the room; their distances (radii) from the array are uniformly sampled between $1$ and $4$~m, and the \ac{DOA} of each source is measured relative to the array center and sampled on a $2^\circ$ grid over $[0^\circ,180^\circ]$.

During training, mixtures and their corresponding targets are generated on the fly. For each mixture, we randomly sample \ac{SNR} levels in the range $0$--$10$~dB and \ac{SIR} levels in the range $5$--$10$~dB. When the textual prompt specifies a loudness difference between the two speakers, the noise signals are normalized with respect to the weaker speaker to preserve the intended loudness ordering in the mixture. In addition to the extraction setting with two speakers and two noises, we also generate an enhancement setting in which only a single randomly chosen speaker is present, mixed with background noise under the same simulation procedure. The dataset consisted of 70,000 training samples, 1,000 validation samples, and 1,000 test samples. \gen{During training, each example is an enhancement example with probability $0.1$ and a two-microphone example with probability $0.5$, and $75\%$ of the two-microphone extraction examples use a spatial prompt.}

\subsection{Implementation Details}
\label{subsec:impl}
\begin{genblock}
All audio signals are sampled at 8 kHz. The \ac{STFT} is computed with 256-sample frames (rectangular window) and a hop size of 128 samples, giving $F=129$ frequency bins. Each encoder contains one TF-Locoformer block, followed by five shared blocks, all with embedding dimension $D=128$; the cross-attention layers use four heads. Text embeddings are generated using the \texttt{e5-small-v2} encoder ($d=384$), which is first contrastively fine-tuned using a triplet margin loss with a margin of 0.5 and then fine-tuned with the extraction network. The complete extraction network contains approximately 52 million parameters, 33 million of which belong to the text encoder. The model is trained using the AdamW optimizer with learning rates of $5\times10^{-5}$ for the acoustic layers, $2\times10^{-4}$ for the conditioning layers, and $1\times10^{-5}$ for the text encoder, all with a 1{,}000-step linear warm-up, and with gradient clipping at 1. Each batch contains 30 mixtures, each used with two prompts (60 examples). Training uses bf16 mixed precision and gradient checkpointing, and an exponential moving average (EMA) of the weights with decay $0.999$ is kept and used for evaluation. The model is trained on a single NVIDIA H200 GPU. 
\end{genblock}

\subsection{Baselines}
We compare our method against established spatial-extraction systems and strong speech-enhancement models. For spatially guided extraction, we compare against SSF, a spatially selective nonlinear filter~\cite{tesch2023multi}, and SpatialNet~\cite{quan2024spatialnet}, a state-of-the-art multichannel network for joint separation, denoising, and dereverberation. \gen{SSF is conditioned on the oracle \ac{DOA} of the target. SpatialNet receives no target cue: it is used unmodified as a two-output separator, trained from scratch on our two-microphone spatial-extraction mixtures with a permutation-invariant SI-SDR loss and with reverberant targets, as for \ours{}. At test time, each prompt is assigned the SpatialNet output given by the oracle permutation, i.e., the assignment of the two outputs to the two reference speakers that maximizes SI-SDR. Both spatial baselines therefore use oracle information that \ours{} does not have.} For speech enhancement, we compare to HDDEMUCS-TF~\cite{HDDEMUCS-TF}\gen{ in the SC setting and to TF-GridNet~\cite{wang2023tf}, a state-of-the-art multi-channel time--frequency network, in the MC setting. TF-GridNet receives both microphones and is trained on our MC enhancement data from scratch, and it is evaluated on the same test items as our model}. For a fair comparison, all models were \gen{trained or fine-tuned} on our dataset.

\subsection{Evaluation Metrics}
We report \gen{four} complementary metrics. \ac{SI-SDR} (in dB) measures the fidelity of the estimated waveform to the reference and is invariant to overall scale. \ac{STOI} estimates speech intelligibility on a $[0,1]$ scale. \gen{DNSMOS P.835~\cite{reddy2022dnsmos} is a non-intrusive estimate of perceived quality; we report its overall score (OVRL) on a $1$--$5$ scale. The word error rate (WER, in \%, lower is better) is obtained with the Whisper large-v3 recognizer~\cite{radford2023robust} after Whisper's English text normalization. The test items are crops of LibriSpeech utterances, so the WER of the mixture and of each output is computed against the Whisper transcript of the reference signal, and it is reported only for targets that contain exactly one speaker (with or without noise). All signals are sampled at 8~kHz and upsampled to 16~kHz for both measures, which lowers the absolute DNSMOS values of all signals, including the clean reference (OVRL of about $2.05$).} For each comparison, we also report the corresponding input (unprocessed mixture) values to explicitly indicate the improvement. \gen{Each test scenario of the extraction tasks is evaluated with both its prompt and its complementary prompt. Every item has a fixed random seed, so all results are computed on identical signals.}

\section{Experimental Results}
\label{sec:results}

\gen{We evaluate the proposed \ours{} framework in five settings: (i) multi-channel (MC) attribute-based extraction, (ii) single-channel (SC) attribute-based extraction, (iii) spatially guided extraction, (iv) MC speech enhancement, and (v) SC speech enhancement. Table~\ref{table:overall} summarizes the overall results, and Tables~\ref{table:mc_extraction} and~\ref{table:sc_extraction} provide per-configuration breakdowns for MC and SC extraction. Table~\ref{table:prompt_type} breaks down attribute-based extraction by prompt formulation: \emph{All} (retain and suppress), \emph{Keep} (retain-only), and \emph{Remove} (suppress-only).}

\subsection{Attribute-Based Extraction}

\gen{As shown in Table~\ref{table:overall}, in MC extraction, SI-SDR improves from approximately $0$~dB at the input to 11.11~dB, with STOI increasing from 0.575 to 0.797. In SC extraction, SI-SDR reaches 10.83~dB, demonstrating that semantic conditioning alone provides strong separation capabilities even in the absence of spatial information. The MC and SC configurations are within $0.3$~dB of each other on attribute-based extraction. The semantic prompts in this task can be resolved from spectral cues alone, and the second microphone contributes mainly to the spatially guided task (Section~\ref{subsec:spatial}).} \gen{For targets that contain a single speaker, the WER drops from about $76$\% at the input to 33.3\% (MC) and 34.4\% (SC). The input WER is high because the recognizer also transcribes the competing speaker.}

\begin{table}[!tbp]
\color{gencolor}
\caption{Overall results across the five settings on the test items. SI-SDR (dB), STOI, and DNSMOS OVRL: higher is better; WER (\%): lower is better. In the attribute-based extraction tasks, WER is computed over the items whose target contains a single speaker.}
\label{table:overall}
\centering
\setlength{\tabcolsep}{3pt}
\resizebox{\columnwidth}{!}{%
\begin{tabular}{@{}lcccccccc@{}}
\toprule
 & \multicolumn{4}{c}{Input} & \multicolumn{4}{c}{\ours{}} \\
\cmidrule(lr){2-5}\cmidrule(l){6-9}
Task & SI-SDR & STOI & DNSMOS & WER & SI-SDR & STOI & DNSMOS & WER \\
\midrule
MC Extraction & -0.03 & 0.575 & 1.54 & 76.1 & 11.11 & 0.797 & 1.61 & 33.3 \\
SC Extraction & -0.03 & 0.575 & 1.53 & 75.8 & 10.83 & 0.789 & 1.61 & 34.4 \\
Spatial Extraction & -3.83 & 0.474 & 1.41 & 100.8 & 10.53 & 0.857 & 2.36 & 29.2 \\
MC Enhancement & 5.70 & 0.792 & 1.71 & 22.1 & 18.66 & 0.960 & 2.30 & 8.2 \\
SC Enhancement & 5.55 & 0.790 & 1.70 & 21.1 & 18.15 & 0.955 & 2.33 & 9.7 \\
\midrule
Average & 1.47 & 0.641 & 1.58 & 59.2 & 13.86 & 0.872 & 2.04 & 22.9 \\
\bottomrule
\end{tabular}}
\end{table}

\gen{The detailed breakdowns in Tables~\ref{table:mc_extraction} and~\ref{table:sc_extraction} reveal how difficulty varies across target configurations. Mixed selections such as \emph{Target + Both Noises} reach 15.27~dB (MC) and 15.32~dB (SC), and extracting both speakers reaches 12.93~dB (MC) and 12.14~dB (SC). In \emph{Single Target}, one speaker must be taken out of a mixture with a second speaker and two noises; the model reaches 11.70~dB (MC) and 11.23~dB (SC), an improvement of about $15$~dB over the input. As expected, extracting a \emph{single} directional noise source from a heavily degraded input (input SI-SDR $\approx -8$~dB) remains the hardest case; nevertheless, the model still improves the input by more than $11$~dB.} \gen{DNSMOS rates any non-speech energy as a degradation, including noise that the prompt asks to keep. It is therefore informative mainly for speech-only targets: for \emph{Single Target} it rises from 1.61 to 2.30 (MC) and 2.27 (SC), and for \emph{Both Targets} from 1.45 to 1.92 (MC) and from 1.42 to 1.88 (SC), whereas outputs that correctly contain the requested noise receive low scores (1.15--1.20 for noise-only targets).}

\begin{table}[!htbp]
\color{gencolor}
\caption{Per-configuration MC extraction results. SI-SDR (dB), STOI, and DNSMOS OVRL: higher is better; WER (\%): lower is better, reported only when the target contains a single speaker.}
\label{table:mc_extraction}
\centering
\setlength{\tabcolsep}{3pt}
\resizebox{\columnwidth}{!}{%
\begin{tabular}{@{}lcccccccc@{}}
\toprule
 & \multicolumn{4}{c}{Input} & \multicolumn{4}{c}{\ours{}} \\
\cmidrule(lr){2-5}\cmidrule(l){6-9}
Action Category & SI-SDR & STOI & DNSMOS & WER & SI-SDR & STOI & DNSMOS & WER \\
\midrule
Both Noises & -3.32 & 0.464 & 1.45 & -- & 9.40 & 0.730 & 1.20 & -- \\
Both Targets & 3.28 & 0.668 & 1.45 & -- & 12.93 & 0.874 & 1.92 & -- \\
Both Targets + Noise & 7.93 & 0.813 & 1.49 & -- & 13.27 & 0.877 & 1.43 & -- \\
Single Noise & -7.95 & 0.282 & 1.49 & -- & 3.99 & 0.491 & 1.15 & -- \\
Single Target & -3.58 & 0.503 & 1.61 & 81.7 & 11.70 & 0.856 & 2.30 & 33.4 \\
Target + Both Noises & 3.48 & 0.709 & 1.61 & 67.6 & 15.27 & 0.917 & 1.53 & 34.8 \\
Target + Noise & -0.03 & 0.580 & 1.57 & 77.6 & 11.47 & 0.833 & 1.68 & 32.4 \\
\bottomrule
\end{tabular}}
\end{table}

\begin{table}[!tbp]
\color{gencolor}
\caption{Per-configuration SC extraction results. SI-SDR (dB), STOI, and DNSMOS OVRL: higher is better; WER (\%): lower is better, reported only when the target contains a single speaker.}
\label{table:sc_extraction}
\centering
\setlength{\tabcolsep}{3pt}
\resizebox{\columnwidth}{!}{%
\begin{tabular}{@{}lcccccccc@{}}
\toprule
 & \multicolumn{4}{c}{Input} & \multicolumn{4}{c}{\ours{}} \\
\cmidrule(lr){2-5}\cmidrule(l){6-9}
Action Category & SI-SDR & STOI & DNSMOS & WER & SI-SDR & STOI & DNSMOS & WER \\
\midrule
Both Noises & -3.33 & 0.462 & 1.42 & -- & 9.01 & 0.723 & 1.18 & -- \\
Both Targets & 3.29 & 0.670 & 1.42 & -- & 12.14 & 0.855 & 1.88 & -- \\
Both Targets + Noise & 7.92 & 0.813 & 1.47 & -- & 13.13 & 0.877 & 1.45 & -- \\
Single Noise & -7.94 & 0.281 & 1.47 & -- & 3.61 & 0.480 & 1.17 & -- \\
Single Target & -3.88 & 0.497 & 1.61 & 81.5 & 11.23 & 0.843 & 2.27 & 36.3 \\
Target + Both Noises & 3.75 & 0.713 & 1.61 & 66.9 & 15.32 & 0.916 & 1.56 & 35.6 \\
Target + Noise & 0.00 & 0.579 & 1.57 & 77.3 & 11.24 & 0.825 & 1.66 & 32.7 \\
\bottomrule
\end{tabular}}
\end{table}

\begin{genblock}
\subsection{Semantic Fidelity of Single-Noise Extraction}
\label{subsec:gemma}

For noise targets, \ac{SI-SDR} is not an adequate measure: it rewards exact waveform reconstruction, whereas the identity of noise-like sound events lies in their spectro-temporal envelope, so perceptually correct outputs can still score low (about 3.8~dB here, from an input of about $-8$~dB). To assess whether the extracted signal is nevertheless the \emph{requested sound event}, we measure its semantic similarity in the joint text--audio embedding space of EmbeddingGemma~2~\cite{google2026embeddinggemma2}, a multimodal embedding model that maps text and audio into a single 768-dimensional space. This model was not used in training. We evaluate the 640 single-noise items (SC and MC extraction test sets, prompt and complementary prompt), and, in addition, the 244 items whose prompt asks for \emph{both} noise sources (\emph{Both Noises}). Each waveform is resampled to 16~kHz and embedded without a prefix, and each noise caption is embedded with the retrieval-query prefix. We report two measures: the cosine similarity between the output embedding and the embedding of the reference noise signal, and the cosine similarity between the output embedding and the caption of the requested noise. For \emph{Both Noises}, the reference is the sum of the two noise signals, and the caption similarity is averaged over the two noise captions. Table~\ref{table:gemma} summarizes the results, together with the unprocessed mixture and the clean reference noise.

We compare with two off-the-shelf language-queried sound separation models, AudioSep~\cite{liu2025audiosep} and CLAPSep~\cite{ma2024clapsep}, using their released pretrained weights without retraining. Both are single-channel models trained at 32~kHz on general sound-event data. They receive the first-microphone mixture resampled to 32~kHz, and their output is resampled back to 8~kHz before scoring. Because these models are trained with short sound-event descriptions rather than instructions, each receives the caption of the requested noise as its text query (for \emph{Both Noises}, the two captions joined by ``and''), whereas \ours{} receives the dataset prompt. CLAPSep also accepts a negative query; using the removed sources as the negative query changed its cosine similarities by at most 0.02, so we report it with the positive query only.

The model moves the output far from the mixture toward the requested noise. In the audio embedding space, the similarity to the reference noise rises from 0.771 for the mixture to 0.950. The caption similarity rises from 0.629 for the mixture to 0.681, which is close to that of the clean reference noise (0.688). The same holds when both noises are requested: the similarity to the reference rises from 0.786 for the mixture to 0.967, and the caption similarity of the output (0.672) is close to that of the clean reference (0.676).

Both language-queried baselines also move the output toward the requested noise, but much less. AudioSep, the stronger of the two, reaches a similarity to the reference of 0.898 for a single noise and 0.908 for both noises, and CLAPSep reaches 0.854 and 0.861, compared with 0.950 and 0.967 for \ours{}. Their caption similarity is also lower (0.677 for AudioSep and 0.660 for CLAPSep for a single noise, versus 0.681). This gap is consistent with a domain mismatch: the test mixtures are reverberant 8~kHz recordings that contain two competing speakers in addition to the two noise sources, a setting that general-purpose sound separators were not trained for.
\end{genblock}

\begin{table}[!tbp]
\color{gencolor}
\caption{Noise extraction on the SC and MC test items (640 single-noise and 244 both-noise items): cosine similarity in the EmbeddingGemma~2 space between the output and the reference noise audio (Ref.) and the requested noise captions (Cap.). Higher is better. AudioSep and CLAPSep use their released weights and receive the requested noise caption as the query.}
\label{table:gemma}
\centering
\normalsize
\setlength{\tabcolsep}{3pt}
\begin{tabular}{@{}lcccc@{}}
\toprule
 & \multicolumn{2}{c}{Single Noise} & \multicolumn{2}{c}{Both Noises} \\
\cmidrule(lr){2-3}\cmidrule(l){4-5}
Signal & Ref. & Cap. & Ref. & Cap. \\
\midrule
Input mixture                       & 0.771 & 0.629 & 0.786 & 0.632 \\
\midrule
AudioSep~\cite{liu2025audiosep}     & 0.898 & 0.677 & 0.908 & 0.671 \\
CLAPSep~\cite{ma2024clapsep}        & 0.854 & 0.660 & 0.861 & 0.652 \\
\ours{}                             & \textbf{0.950} & \textbf{0.681} & \textbf{0.967} & \textbf{0.672} \\
\midrule
Reference noise                     & 1.000 & 0.688 & 1.000 & 0.676 \\
\bottomrule
\end{tabular}
\end{table}

\subsection{Spatially Guided Extraction}
\label{subsec:spatial}

\gen{Table~\ref{table:spatial_extraction} presents a direct comparison of the proposed methodology with SSF and SpatialNet on the spatial extraction task. Although \ours{} is designed as a general text-guided extraction framework rather than a spatially specialized separator, it consistently outperforms the evaluated spatial baselines in both SI-SDR and STOI across all reported spatial conditions.}

\begin{figure}[!tbp]
    \centering
    \includegraphics[width=\columnwidth]{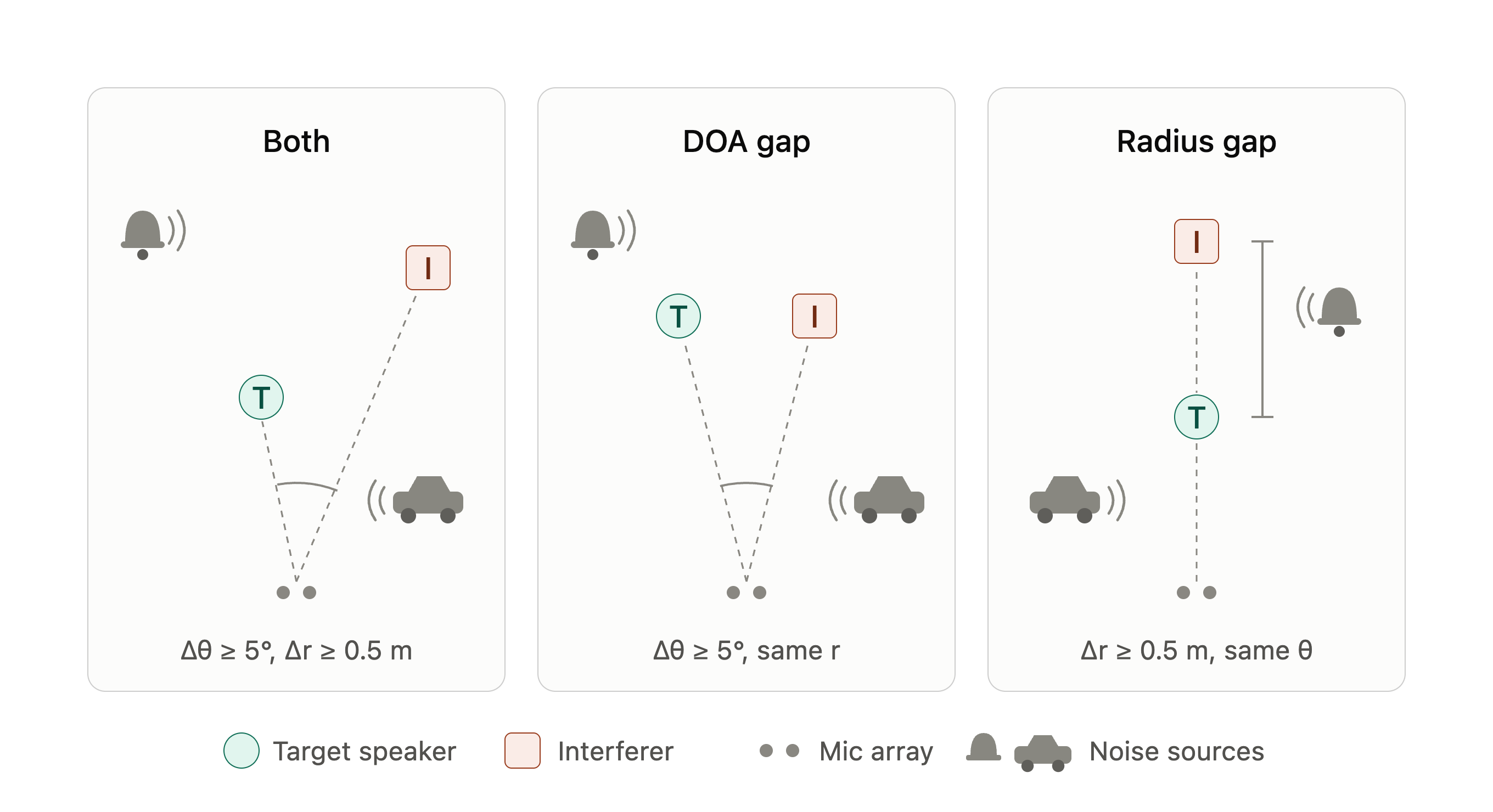}
    \caption{\gen{The three spatial configurations of the spatially guided extraction test set. The target speaker (T) and the interfering speaker (I) differ in both \ac{DOA} and distance (``Both''), only in \ac{DOA} (``DOA gap''), or only in distance from the two-microphone array (``Radius gap''). Two directional noise sources are present in every scene.}}
    \label{fig:spatial_example}
\end{figure}

\noteb{We evaluate spatially guided extraction under three spatial configurations, depending on the relative position of the target and interfering speakers. ``Both'' denotes scenarios in which the speakers differ in both \ac{DOA} and radial distance, with gaps of at least $5^\circ$ and $0.5$~m, respectively. ``DOA gap'' and ``Radius gap'' denote scenarios in which the speakers differ only in azimuth by at least $5^\circ$ or only in radial distance by at least $0.5$~m, respectively. Fig.~\ref{fig:spatial_example} illustrates the three configurations.}
 \gen{Under joint azimuth and distance discrimination (``Both''), \ours{} reaches 10.67~dB, compared with $5.70$~dB for SpatialNet and $5.00$~dB for SSF. In the azimuth-only condition (``DOA gap''), it again achieves the strongest separation among the evaluated systems (10.41~dB). In the distance-only scenario (``Radius gap''), which is the most challenging condition for the baselines, \ours{} reaches 10.52~dB, more than $6.5$~dB above SpatialNet. SSF is not evaluated in this condition: it is conditioned on the \ac{DOA} of the target, so when the two speakers share the same \ac{DOA} and differ only in distance, its cue cannot identify the target. The performance of \ours{} is nearly uniform across the three conditions, so distance-based prompts are resolved about as well as azimuth-based prompts. In all three conditions, STOI reaches about $0.86$, compared with $0.69$--$0.71$ for SpatialNet and $0.65$--$0.69$ for SSF. The output is also more intelligible and of higher perceived quality: its WER is 28.0--31.6\%, compared with 59.2--63.9\% for SpatialNet and 57.4--66.2\% for SSF, and its DNSMOS is 2.32--2.41, compared with 1.96--1.98 and 1.75--1.79, respectively. The input WER exceeds 100\% because the recognizer transcribes both speakers.} Importantly, whereas SSF and SpatialNet are spatially specialized architectures \gen{that rely on oracle information (SSF on the target \ac{DOA}, SpatialNet on the oracle assignment of its two outputs), so that neither has to identify the target itself}, \ours{} operates within a unified SC--MC framework, supports free-form natural-language spatial descriptions, and simultaneously enables semantic attribute-based conditioning. Thus, the proposed approach not only achieves superior spatial extraction performance compared with the other methods but also offers substantially greater flexibility.

\begin{table}[!tbp]
\color{gencolor}
\caption{Spatially guided extraction results. SI-SDR (dB), STOI, and DNSMOS OVRL: higher is better; WER (\%): lower is better. SSF is conditioned on the oracle \ac{DOA} of the target; it has no ``Radius gap'' row because a \ac{DOA} cue cannot distinguish speakers that differ only in distance.\gen{ SpatialNet separates both speakers without a cue, and its outputs are assigned to the prompts by the oracle permutation.}}
\label{table:spatial_extraction}
\centering
\normalsize
\setlength{\tabcolsep}{4pt}
\resizebox{\columnwidth}{!}{%
\begin{tabular}{@{}llcccc@{}}
\toprule
Method & Action Category & SI-SDR $\uparrow$ & STOI $\uparrow$ & DNSMOS $\uparrow$ & WER $\downarrow$ \\
\midrule
\multirow{3}{*}{Input}
 & Both       & -3.99 & 0.466 & 1.42 & 101.3 \\
 & DOA gap    & -3.75 & 0.475 & 1.41 & 104.3 \\
 & Radius gap & -3.75 & 0.480 & 1.41 & 96.7 \\
\midrule
\multirow{2}{*}{SSF}
 & Both       & 5.00 & 0.689 & 1.75 & 66.2 \\
 & DOA gap    & 3.87 & 0.650 & 1.79 & 57.4 \\
\midrule
\multirow{3}{*}{SpatialNet}
 & Both       & 5.70 & 0.71 & 1.98 & 62.4 \\
 & DOA gap    & 5.70 & 0.71 & 1.96 & 59.2 \\
 & Radius gap & 3.80 & 0.69 & 1.96 & 63.9 \\
\midrule
\multirow{3}{*}{\ours{}}
 & Both       & \textbf{10.67} & \textbf{0.856} & \textbf{2.36} & \textbf{31.6} \\
 & DOA gap    & \textbf{10.41} & \textbf{0.856} & \textbf{2.32} & \textbf{28.0} \\
 & Radius gap & \textbf{10.52} & \textbf{0.857} & \textbf{2.41} & \textbf{28.1} \\
\bottomrule
\end{tabular}}
\end{table}

\begin{figure}[!tbp]
    \centering
    \includegraphics[width=\columnwidth]{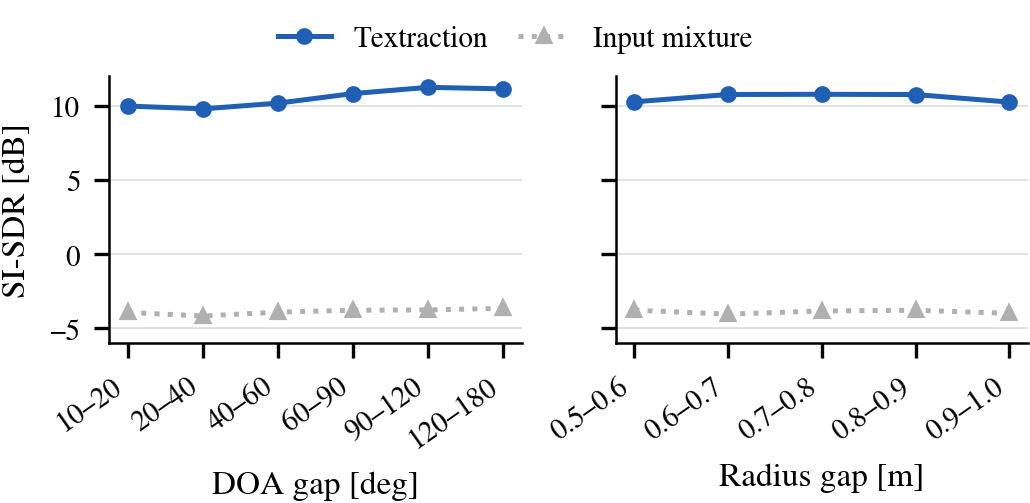}
    \caption{\gen{Spatially guided extraction SI-SDR as a function of the spatial separation between the two speakers. Left: \ac{DOA} gap, over the scenes in which the speakers differ in \ac{DOA} (``Both'' and ``DOA gap''); right: radius gap, over the scenes in which they differ in distance (``Both'' and ``Radius gap''). Each point is the mean over the test items (prompt and complementary prompt) in the bin.}}
    \label{fig:spatial_gaps}
\end{figure}

\gen{Fig.~\ref{fig:spatial_gaps} breaks the spatial results down by how far apart the two speakers are. In \ac{DOA}, \ours{} reaches 9.99~dB SI-SDR for gaps of only $10^\circ$--$20^\circ$ and 10.8--11.3~dB for gaps beyond $60^\circ$, so speakers that are close in direction remain separable, and a wider angular separation adds only about 1~dB. In distance, \ours{} is nearly flat at 10.3--10.8~dB over the whole range of radius gaps ($0.5$--$1.0$~m), so prompts that refer to a closer or farther speaker are resolved as reliably for a $0.5$~m difference as for a $1$~m one. }

\subsection{Speech Enhancement}

\gen{In the enhancement task (Tables~\ref{table:mc_enh} and~\ref{table:sc_enh}), \ours{} achieves 18.66~dB SI-SDR / 0.960 STOI in MC and 18.15~dB SI-SDR / 0.955 STOI in SC. In the MC setting, the dedicated TF-GridNet reaches $18.15$~dB SI-SDR / $0.959$ STOI, so \ours{} surpasses a state-of-the-art enhancement network trained for this task alone by $0.5$~dB SI-SDR, with a similar STOI. In the SC setting, \ours{} surpasses HDDEMUCS-TF ($13.72$~dB SI-SDR / $0.890$ STOI) by about $4.4$~dB in SI-SDR and $0.065$ in STOI. In MC enhancement, the DNSMOS of \ours{} is slightly higher than that of TF-GridNet (2.30 vs.\ 2.24), and the two reach a similar WER (8.2\% vs.\ 7.9\%), both far below the 22.1\% of the input. In SC enhancement, \ours{} reduces the WER from 21.1\% to 9.7\%, whereas the output of HDDEMUCS-TF has a higher WER than the input (25.9\%), and \ours{} also reaches a higher DNSMOS (2.33 vs.\ 2.24). This is notable because HDDEMUCS-TF and TF-GridNet are dedicated enhancement models, whereas \ours{} performs enhancement as a special case of a more general text-guided extraction framework, with the same weights that also serve the four extraction tasks.}

\begin{table}[!htbp]
\color{gencolor}
\caption{MC enhancement results on the same test items. WER in \%.}
\label{table:mc_enh}
\centering
\resizebox{\columnwidth}{!}{%
\begin{tabular}{lcccc}
\toprule
Method & SI-SDR $\uparrow$ & STOI $\uparrow$ & DNSMOS $\uparrow$ & WER $\downarrow$ \\
\midrule
Input            & 5.70 & 0.792 & 1.71 & 22.1 \\
\midrule
TF-GridNet       & 18.15 & 0.959 & 2.24 & \textbf{7.9} \\
\ours{}          & \textbf{18.66} & \textbf{0.960} & \textbf{2.30} & 8.2 \\
\bottomrule
\end{tabular}}
\end{table}

\begin{table}[!htbp]
\color{gencolor}
\caption{SC enhancement results. WER in \%.}
\label{table:sc_enh}
\centering
\resizebox{\columnwidth}{!}{%
\begin{tabular}{lcccc}
\toprule
Method & SI-SDR $\uparrow$ & STOI $\uparrow$ & DNSMOS $\uparrow$ & WER $\downarrow$ \\
\midrule
Input            & 5.55 & 0.790 & 1.70 & 21.1 \\
\midrule
HDDEMUCS-TF      & 13.72 & 0.890 & 2.24 & 25.9 \\
\ours{}          & \textbf{18.15} & \textbf{0.955} & \textbf{2.33} & \textbf{9.7} \\
\bottomrule
\end{tabular}}
\end{table}

\subsection{Robustness to Prompt Formulation}

\gen{Table~\ref{table:prompt_type} groups the attribute-based extraction test items by the formulation of their prompt. The performance differences between ``All,'' ``Keep,'' and ``Remove'' prompts are small: SI-SDR differences are within $0.55$~dB, STOI differences are within $0.016$, \gen{and the WER ranges from 32.1\% to 36.7\%}. ``All'' prompts, which contain both retain and suppress clauses, are slightly harder than the other two formulations. This confirms that the fine-tuned E5 encoder effectively captures instruction polarity, that the model is robust to linguistic variability and paraphrasing, and that both positive (retain) and negative (suppress) instructions are interpreted consistently.}

\begin{table}[!tbp]
\color{gencolor}
\caption{Attribute-based extraction by prompt formulation. WER is computed over the items whose target contains a single speaker.}
\label{table:prompt_type}
\centering
\normalsize
\setlength{\tabcolsep}{4pt}
\resizebox{\columnwidth}{!}{%
\begin{tabular}{@{}llrcccc@{}}
\toprule
Task & Prompt & Items & SI-SDR $\uparrow$ & STOI $\uparrow$ & DNSMOS $\uparrow$ & WER $\downarrow$ \\
\midrule
MC Extraction & All & 620 & 10.83 & 0.792 & \textbf{1.63} & 35.6 \\
 & Keep & 708 & 11.11 & \textbf{0.800} & 1.61 & 32.3 \\
 & Remove & 670 & \textbf{11.37} & 0.799 & 1.58 & \textbf{32.1} \\
\midrule
SC Extraction & All & 667 & 10.63 & 0.780 & \textbf{1.61} & 36.7 \\
 & Keep & 642 & \textbf{11.11} & \textbf{0.796} & \textbf{1.61} & \textbf{33.2} \\
 & Remove & 689 & 10.77 & 0.792 & 1.60 & 33.4 \\
\bottomrule
\end{tabular}}
\end{table}

\gen{Each extraction test mixture is evaluated with both its prompt and its complementary prompt, which describe different targets in the same mixture. To verify that the output is actually controlled by the text and not by an acoustic prior (e.g., ``return the loudest speaker''), we swapped the prompts of each validation pair, so that every mixture was processed with the prompt of the complementary target. The SI-SDR with respect to the original target then dropped to $-23.0$~dB (SC extraction), $-22.4$~dB (MC extraction), and $-33.5$~dB (spatial extraction), which shows that the model switches its output to the other sound.} \gen{Appendix~\ref{app:typos} evaluates the robustness to typos in the prompt.}

\subsection{Discussion}

\gen{The results demonstrate that \ours{} consistently outperforms the evaluated spatial baselines while simultaneously supporting semantic and spatial text conditioning within a unified SC--MC architecture. In addition, it surpasses the dedicated SC enhancement baseline by a wide margin and matches or outperforms the dedicated MC enhancement network TF-GridNet. These results come from three design choices: a single network in which the prompt conditions every block, so that no separation or selection stage can fail; dedicated SC and MC encoders together with an SC/MC embedding, which let the shared blocks exploit spatial cues whenever a microphone array is available; and a strong time--frequency backbone initialized from a source separator. Together with its robustness to prompt formulation, these findings highlight text-guided conditioning as a flexible and competitive paradigm for speaker extraction and enhancement, capable of subsuming several specialized tasks in a single model.}

\section{Limitations and Future Work}
\label{sec:limitations}

Our study has several limitations that also point to promising directions for future work. First, all experiments are conducted on \emph{simulated} reverberant mixtures; evaluating on real recordings captured by microphone arrays in reverberant acoustic environments is an important next step. Second, the extraction setting assumes at most two simultaneous speakers and two directional noise sources; scaling to more sources and to a variable, unknown number of entities remains open. Third, single-directional noise extraction from strongly degraded inputs remains challenging \gen{(about 3.8~dB SI-SDR)}, suggesting a need for stronger semantic grounding of noise events. Fourth, the spatial descriptions are drawn from a fixed set of relative relations; supporting richer, quantitative spatial language (e.g., approximate angles or distances) is a natural extension. \gen{Fifth, the TF-Locoformer backbone operates on every time--frequency bin and is relatively slow; a lighter backbone would reduce the inference cost.} Finally, while our lightweight text encoder is efficient and polarity-aware, jointly modeling more compositional prompts (nested keep/remove clauses, counting, ordering) is an interesting future research avenue.

\section{Conclusion}
\label{sec:conclusion}

We introduced \textsc{Textraction}, a unified text-guided framework for speaker extraction and speech enhancement that operates across single- and multi-channel inputs within a single model. \gen{Dedicated SC and MC encoders feed shared TF-Locoformer blocks that are conditioned on the prompt and on the recording configuration in every block, and the network estimates the target in one forward pass, with no separation or selection stage.} The framework accepts free-form natural-language prompts that specify targets by semantic attributes, by relative spatial relations, or by both, and that instruct the model to keep, remove, or jointly keep-and-remove arbitrary subsets of speakers and directional noise sources. Rather than relying on a large language model, we obtain conditioning embeddings from a lightweight text-retrieval encoder that is contrastively fine-tuned to be sensitive to instruction polarity while robust to paraphrasing. Across attribute-based extraction, spatially guided extraction, and speech enhancement, \textsc{Textraction} isolates the desired entities under diverse acoustic and spatial conditions, exploits spatial information when a microphone array is available, and matches or surpasses dedicated spatial-extraction and enhancement baselines---while being markedly more flexible. These results position text-guided conditioning as a practical and competitive interface for controllable source extraction, and motivate future work on real-array evaluation, larger and variable source counts, and richer compositional prompting.
\appendices
\section{Robustness to Typos}
\label{app:typos}

\begin{figure}[!tbp]
    \centering
    \includegraphics[width=\columnwidth]{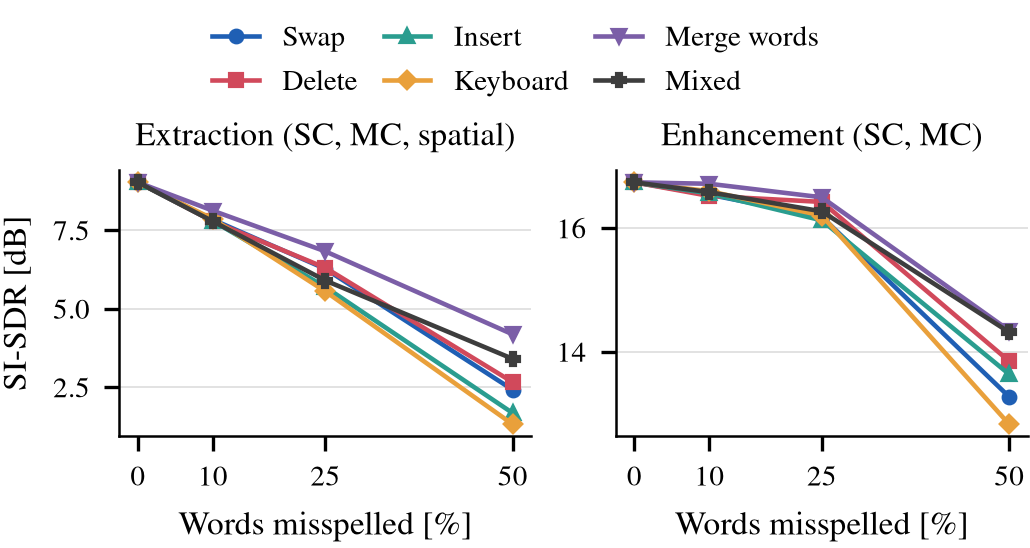}
    \caption{\gen{SI-SDR as a function of the share of misspelled words in the prompt, per typo type. Left: mean over SC, MC, and spatially guided extraction; right: mean over SC and MC enhancement. 0\% is the clean dataset prompt.}}
    \label{fig:typo_curves}
\end{figure}

\begin{genblock}
Prompts typed by users contain typos. To measure how \ours{} handles them, we take 200 test scenes per task, each with two prompts (400 items per task, 2{,}000 in total), and misspell the dataset prompt in five ways. \emph{Swap} exchanges two adjacent letters, \emph{Delete} drops a letter, \emph{Insert} adds a letter from a neighboring key, \emph{Keyboard} replaces a letter with a neighboring key on a QWERTY keyboard, and \emph{Merge words} removes the space to the next word. \emph{Mixed} picks one of these five per word. Each type is applied to 10\%, 25\%, or 50\% of the words of at least three letters. Two targeted conditions misspell every \emph{instruction} word (e.g., \emph{keep}, \emph{remove}, \emph{extract}, \emph{isolate}) or every \emph{description} word (all other words, which describe the sources), with one letter-level typo per word. Table~\ref{tab:typo_examples} shows one prompt under every condition. Because this evaluation uses a 200-scene subset of each task's test set, the clean-prompt values in Table~\ref{tab:typo_results} differ slightly from the full-test-set values in Table~\ref{table:overall}; all conditions in this appendix are evaluated on the same subset and are directly comparable. Besides SI-SDR and STOI, we report for the three extraction tasks how often the output \emph{switched} to the complementary sound: the share of outputs whose SI-SDR with respect to the target of the complementary prompt of the same mixture exceeds the SI-SDR with respect to their own target.

Fig.~\ref{fig:typo_curves} shows SI-SDR as the share of misspelled words grows. Enhancement is almost unaffected up to 25\% of the words (16.36--17.05~dB, compared with 17.08~dB for the clean prompt), because its prompts describe the same target in every scene. Extraction degrades gradually: from 9.48~dB for the clean prompt to 7.99--8.57~dB with 10\% of the words misspelled and 5.59--7.17~dB with 25\% (Table~\ref{tab:typo_results}), and the share of switched outputs rises from 0.8\% to 7.2--11.7\% at 25\%. With half of the words misspelled, extraction falls to 1.11--4.04~dB and 16.0--27.8\% of the outputs switch. Merged words are the mildest typo, since both halves of a merged word remain intact, whereas \emph{Keyboard} and \emph{Insert}, which introduce letters that are not in the word, are the most harmful.

The \emph{Instruction words} and \emph{Description words} rows of Table~\ref{tab:typo_results} show which words matter. A typo in every instruction word (2.2 words per prompt on average) costs 7.23~dB in SC extraction and 6.40~dB in MC extraction, and 12.9\% of the extraction outputs switch to the complementary sound, because the instruction words carry the polarity (\emph{keep} versus \emph{remove}) of attribute-based prompts. In spatially guided extraction and in enhancement, the instruction word does not change the target, and the same typos cost at most 0.79~dB. Typos in every description word (8.8 words per prompt) are the most harmful condition: extraction drops by 10.0--12.3~dB and 36.0\% of the outputs switch, while enhancement loses about 2~dB. Misspelled descriptions remove the information that identifies the target, so that the model can no longer tell which source is requested.
\end{genblock}

\begin{table}[!tbp]
\color{gencolor}
\caption{SI-SDR (dB) and STOI with 25\% of the words misspelled, per typo type and task, and with every instruction word or every description word misspelled. Avg.: mean over the five tasks.}
\label{tab:typo_results}
\centering
\resizebox{\columnwidth}{!}{%
\begin{tabular}{@{}lcccccc@{}}
\toprule
 & \multicolumn{3}{c}{Extraction} & \multicolumn{2}{c}{Enhancement} & \\
\cmidrule(lr){2-4}\cmidrule(lr){5-6}
Prompt & SC & MC & Spatial & SC & MC & Avg. \\
\midrule
\multicolumn{7}{@{}l}{\textit{SI-SDR (dB)}} \\
Clean prompt & 9.93 & 9.84 & 8.67 & 16.90 & 17.26 & 12.52 \\
\midrule
Swap & 6.70 & 6.08 & 6.28 & 16.25 & 16.62 & 10.38 \\
Delete & 5.94 & 6.47 & 6.21 & 16.49 & 16.92 & 10.41 \\
Insert & 5.86 & 5.72 & 6.10 & 16.18 & 16.54 & 10.08 \\
Keyboard & 5.85 & 5.25 & 5.67 & 16.21 & 16.58 & 9.91 \\
Merge words & 7.32 & 6.87 & 7.31 & 16.74 & 17.11 & 11.07 \\
Mixed & 6.15 & 5.38 & 6.29 & 16.35 & 16.69 & 10.17 \\
\midrule
Instruction words & 2.70 & 3.44 & 8.46 & 16.39 & 16.46 & 9.49 \\
Description words & -2.06 & -2.43 & -1.32 & 15.24 & 15.25 & 4.94 \\
\midrule
\multicolumn{7}{@{}l}{\textit{STOI}} \\
Clean prompt & 0.764 & 0.767 & 0.799 & 0.944 & 0.945 & 0.844 \\
\midrule
Swap & 0.688 & 0.680 & 0.748 & 0.936 & 0.936 & 0.798 \\
Delete & 0.677 & 0.692 & 0.747 & 0.937 & 0.940 & 0.799 \\
Insert & 0.679 & 0.672 & 0.743 & 0.935 & 0.935 & 0.793 \\
Keyboard & 0.676 & 0.667 & 0.734 & 0.935 & 0.937 & 0.790 \\
Merge words & 0.701 & 0.693 & 0.770 & 0.941 & 0.943 & 0.810 \\
Mixed & 0.685 & 0.671 & 0.746 & 0.936 & 0.937 & 0.795 \\
\midrule
Instruction words & 0.601 & 0.621 & 0.795 & 0.935 & 0.933 & 0.777 \\
Description words & 0.488 & 0.480 & 0.576 & 0.921 & 0.920 & 0.677 \\
\bottomrule
\end{tabular}}
\end{table}

\begin{table*}[!tbp]
\color{gencolor}
\caption{One MC-extraction test prompt under every typo condition (25\% of the words for the six typo types).}
\label{tab:typo_examples}
\centering
\small
\begin{tabular}{@{}lp{0.80\textwidth}@{}}
\toprule
Condition & Prompt \\
\midrule
Clean prompt & Maintain the loud female speaker, include the bell ringing sound, keep the soft male speaker, and exclude the running motor vehicle engine. \\
Swap & Maitnain the loud female spaeker, include the bell rniging sound, keep the sotf male speaker, adn exclude the rnuning motor vehicle engine. \\
Delete & Mintain te loud fmale speaker, include the bell ringing sound, keep the soft male speaker, and exclude the running mtor vehcle engne. \\
Insert & Maintain thue loud femalpe speakoer, incluide the bell ringing sokund, keep the soft male speaker, and exzclude the running motor vehicle engine. \\
Keyboard & Maintain the loud female speaker, incluce the bell ribging sound, keep the soft male speqker, amd exclude tge running motor vdhicle engine. \\
Merge words & Maintainthe loud femalespeaker, includethe bell ringingsound, keep the soft male speaker, and exclude the running motor vehicle engine. \\
Mixed & Maintain the loud ffmale speaker, include tme bell ringing sound, keepthe soft mael speaker, and ecclude the running motor vehicle engine. \\
Instruction words & Mainain the loud female speaker, inlude the bell ringing sound, kefep the soft male speaker, and exlude the running motor vehicle engine. \\
Description words & Maintain thge lpud femle spaker, include teh belpl ringign soud, keep tge sft mawle sleaker, angd exclude tge runnign moitor vehikcle enginhe. \\
\bottomrule
\end{tabular}
\end{table*}

\bibliographystyle{IEEEtran}
\bibliography{main}

\end{document}